\documentclass[smallextended]{svjour3}

\smartqed

\usepackage{graphicx}
\usepackage{amssymb}
\usepackage{epsfig}
\usepackage{amsmath}
\usepackage{float}
\usepackage{color}
\usepackage[authoryear]{natbib}
\usepackage{appendix}

\newcommand{\be}{\begin{equation}}
\newcommand{\ee}{\end{equation}}
\newcommand{\bea}{\begin{eqnarray}}
\newcommand{\eea}{\end{eqnarray}}

\newcommand{\nd}{\noindent}

\begin{document}

\title{Modeling Age-Dependent Radiation-Induced Second Cancer Risks and Estimation of Mutation Rate: An Evolutionary Approach}

\titlerunning{Modeling Age Dependent Radiation Induced Second Cancer}      

\author{Kamran Kaveh \and Venkata S. K. Manem \and Mohammad Kohandel \and Siv Sivaloganathan}
\institute{K.Kaveh, V. S. K. Manem \at Department of Applied Mathematics, University of Waterloo,\\ Waterloo, ON, Canada N2L 3G1 \and K.Kaveh \at email: kkavehma@gmail.com\and K. Kohandel, S. Sivaloganathan \at 1. Department of Applied Mathematics, University of Waterloo,\\ Waterloo, ON, Canada N2L 3G1\\ 2. Center for Mathematical Medicine,\\ Fields Institute for Research in Mathematical Sciences,\\ Toronto, ON, Canada M5T 3J1}

\maketitle

\begin{abstract}
Although the survival rate of cancer patients has significantly increased due to advances in anti-cancer therapeutics, one of the major side effects of these therapies, particularly radiotherapy, is the potential manifestation of radiation-induced secondary malignancies. 
In this work, a novel evolutionary stochastic model is introduced that couples short-term formalism (during radiotherapy) and 
long-term formalism (post treatment). This framework is used to estimate the risks of second cancer as a function of spontaneous 
background and radiation-induced mutation rates of normal and pre-malignant cells. By fitting the model to available clinical data for 
spontaneous background risk together with data of Hodgkin's lymphoma survivors (for various organs), the second cancer mutation rate is estimated. 
The model predicts a significant increase in mutation rate for some cancer types, which may be a sign of genomic instability. Finally, it is shown
that the model results are in agreement with the measured results for excess relative risk (ERR) as a function of exposure age, and 
that the model predicts a negative correlation of ERR with increase in attained age. This novel approach can be used to analyze several 
radiotherapy protocols in current clinical practice, and to forecast the second cancer risks over time for individual patients.

\end{abstract}


\section{Introduction}

The evolution of radiation medicine coupled with various technological advances, in therapeutic radiation treatments has dramatically improved 
the survival rate of cancer patients over the last several decades. Current treatment techniques deliver doses to the gross tumor volume with 
great precision. However this comes at the cost of damaging healthy tissue in the vicinity of the tumor, leading to the possible
manifestation of a secondary tumor post treatment in many cancer survivors, which may occur sometimes several decades after treatment\citep{xu2008review}. 
Due to the higher survival rate of cancer patients in recent years, the investigation of secondary malignancy risks due to radiation therapy 
has now become of greater importance than ever. The fact that the survival rate of patients (for example Hodgkin's Lymphoma, Leukemia and Sarcomas) has improved to several more decades suggests that one needs to construct a model that can predict the age and time-dependent risks of second cancers. Radiation-induced cancer risks have been well documented in atomic bomb survivors mortality studies \citep{key:preston}\citep{key:little}. 
There is also evidence from epidemiological studies that the mortality rate of Hodgkin's Lymphoma survivors is high post radiation treatment 
\citep{key:AVK}\citep{key:BYR}. Several cohort and case control studies have indicated that there is an increased risk of second malignancies 
with young Hodgkin's Lymphoma survivors post irradiation \citep{key:bhatia}\citep{key:hodgson1}. Recently, several clinical and modeling studies have described 
the risk of of second malignancies for various neighboring organs \citep{key:sachs}\citep{xu2008review}\citep{moteabbed2014risk}\citep{paganetti2012assessment}\citep{athar2011comparison}\citep{hall2006intensity}\citep{manem2014radquality}. Some of these clinical studies have suggested that there is 
a high incidence rate of breast cancer among young Hodgkin's Lymphoma survivors compared to women treated in middle age \citep{key:TD}. 
Children treated with radiotherapy are particularly vulnerable to radiation-induced carcinogenesis \citep{key:bhatia}. Several clinical 
findings indicate that second cancer risks change with respect to age at exposure and also with respect to time since exposure 
\citep{key:TD}\citep{key:DMC}. Some authors studied the efficacy of dose escalation on secondary cancer induction\citep{manem2014efficacy}\citep{schneider2007impact}.
It has also been noted that the latency period is typically between $10$ to $20$ years for solid tumors, and around $5$ years for leukemia 
post irradiation \citep{zhang2012risk}\citep{wang2014second}\citep{schneider2014radiation}\citep{key:VKS}\citep{key:yeoh}\citep{key:yahalom}\citep{key:hodgson2007long}. As a result of increasing prevalence of secondary malignancies, it is clearly of great importance to 
investigate the risks associated with different treatment regimens as well as their possible variation with age (by incorporating 
age at exposure and time since exposure)\citep{key:hodgson1}.
Site-specific dose-response relationships for cancer induction from the combined Japanese A-bomb and Hodgkin cohorts for doses on a par with 
radiation bomb survivors, has been actively studied in the literature \citep{key:preston}\citep{key:preston4}. The approaches taken are 
phenomenological models assuming the temporal pattern of risk to be an exponential power law \citep{key:preston}. With respect to the 
baseline rates, \citep{key:preston} showed that these are proportional to the attained age raised to the power of $5.5$, and decrease 
around the age of $90$. Another point to take into consideration is that gender appears to play a major role in defining the baseline rates. 
This type of analysis of cancer incidence among A-bomb survivors indicates that the cancer risk (excess relative risk) is a function of time 
and decreases with age, thus leading to the conclusion that, for long-term age dependent models, both exposure age and attained age are important 
factors to incorporate. At the same time however, age-dependent factors are difficult to model in practice.
Radiation-induced second cancers are thought to arise as a result of multiple mutations of healthy cells (or pre-cancerous, 
pre-malignant cells) post irradiation of the primary tumor \citep{key:sachs}\citep{lindsay2001radiation}\citep{schneider2009mechanistic}. Pre-malignant cells are those cells that have the potential to turn into malignant cells. 
The process is very complex to model due to variations in various biological mechanisms in patients (that depend on habits, or environmental exposure). 
As discussed in the literature on primary cancer age incidence modeling \citep{key:armitage}\citep{key:doll}, the manifestation of 
a second cancer is considered here to be the result of a two-hit process. It is assumed that anti-cancer therapies, such as 
radiation treatment, introduce a first hit to some of the normal cells in the vicinity of the tumor. This introduces a sub-population of 
radiation-induced pre-malignant cells in the system (see \citep{key:sachs} and references therein). The detailed bio-physical mechanisms of pre-malignant cells 
resulting from radiation exposure, are still not well understood. Additionally, radiation therapy has a destructive effect on the tissue 
architecture, leading to some changes in the spatial structure of the tissue (see for example \citep{key:tissuearch}). It has also been suggested 
that radiation therapy can introduce chromosomal instabilities, and thus make the second hit mutations more effective and unpredictable in 
nature \citep{key:radind-cin}. In the presence of genomic instability, the possibility of having a hit on a particular locus is broadened into a 
range of hits and changes across a pair of chromosomes. However, there is no direct evidence for genomic instability as a result of radiation 
exposure, and there are several ongoing debates on some intriguing questions related to the conditions required for genomic instability 
\citep{key:bhatia}\citep{key:alice}.
On the other hand, the progression of cancer and evolutionary dynamics of two-stage models of cancer progression have been discussed in the 
existing literature. The latter were developed to quantify the dynamics of how an advantageous mutant (pre-malignant cell) can outgrow a 
background population of normal stem cells and upon a second (malignant) mutation resulting in 
a malignant niche and eventually resulting to a secondary malignancy. 
These models have also been 
successfully applied to describe dynamics of chromosomal instability by different authors \citep{key:nowak-main}\citep{key:nowak-cin}. 
It has also been proposed that second cancers may arise due to the creation of residual pre-malignant cells (cells due to radiation treatment) 
which can proliferate in between therapies and eventually grow afterwards \citep{key:sachs}.
In this paper, a mathematical model is construced to predict the excess relative risk (ERR) of second cancers due to radiation exposure 
during treatment of a primary malignancy. The presented framework  captures the stochastic nature of such complex evolutionary systems. 
The parameters of the model are: (i) the mutation rate after radiation exposure which is assumed to be different from the background value 
of the mutation rate and, (ii) the proliferation strength of the pre-malignant cells. The dependence of the ERR on attained age and age at exposure 
of the patients is predicted by the model.

\section{Materials and Methods}

In this section the mathematical model used to estimate the ERR due to radiation exposure is briefly discussed. The mathematical details of the 
model are presented in Appendix A. The framework used is built on the evolutionary model of cancer initiation/progression of 
two-hit mutation models of tumor suppressor gene inactivation (TSG), and the evolutionary dynamics of chromosomal instability
developed by Nowak et al. \citep{key:nowak-main}\citep{key:nowak-cin}.

In modelling the risks of secondary malignancies post irradiation, a new time scale  (age at exposure $t_{e}$) is introduced to capture the 
effect of radiation-induced pre-malignant cell initiation. It is assumed that radiation-induced secondary cancers are initiated 
by mutated stem cells. In a two-hit mechanism, it is further assumed that normal stem cells can be mutated into pre-malignant cells 
(also known as type-1, or one-hit cells). This random event occurs either by spontaneous mutation, or, as a result of a radiation-induced 
initiation mechanism during treatment. Each pre-malignant cell can then undergo a second spontaneous mutation and transform into a malignant 
phenotype (also known as a type-2, or two-hit cell). Malignant cells can eventually initiate tumor clones. The two hit hypothesis here is a 
reflection of the fact that one is focusing on the malignancies induced by tumor suppressor gene inactivation (TSG). 
In the case of TSG inactivation, a single mutation does not lead to loss-of-function for a cell, due to which it is assumed that 
pre-malignant cells have a very weak selection advantage/disadvantage over normal stem cells; while malignant cells are expected to have a 
significantly higher proliferation potential.

The premise is that stem cells populate stem cell niche areas in the tissue which are homeostatically regulated. The proliferation/selection 
of the three types of stem cells- normal, pre-malignant, malignant- determine the phenotype that will eventually take over the corresponding niche. 
It is assumed that the selection mechanism is governed by a stochastic birth-death model, (i.e. Moran process \citep{key:M}\citep{key:durrett}). 
The form of the growth term of advantaged phenotypes (normally type-2 and also marginally type-1 cells) is reminiscent of logistic growth with the 
carrying capacity being the niche size. It is assumed that the total number of stem cells (either normal, pre-malignant, or malignant phenotypes)
 is a constant, $N$ (within each niche). This number can vary between $10-10^{6}$ (depending on the organ type). The total number of the niches 
is denoted by $\tilde{N}$, which is generally a large number ($\sim 10^{7}$). The first and second spontaneous (background) mutation rates 
are denoted by $u_{1}$ and $u_{2}$. The second mutation rate for the irradiated (normal) tissues is distinguished by $u_{r}$ 
(rather than $u_{2}$) to indicate the secondary effects of radiation on pre-malignant cells. One of the probable secondary effects of ionizing 
radiation may be radiation-induced genomic instability which has been suggested in the literature 
(\citep{key:alice}, \citep{key:nowak-main}, \citep{key:nowak-cin}). Moreover, the number of {\it malignant} stem cells created by radiation 
is ignored due to their small number relative to the pre-malignant cells.\newline

\begin{figure}[!h]
\begin{center}
\epsfig{figure=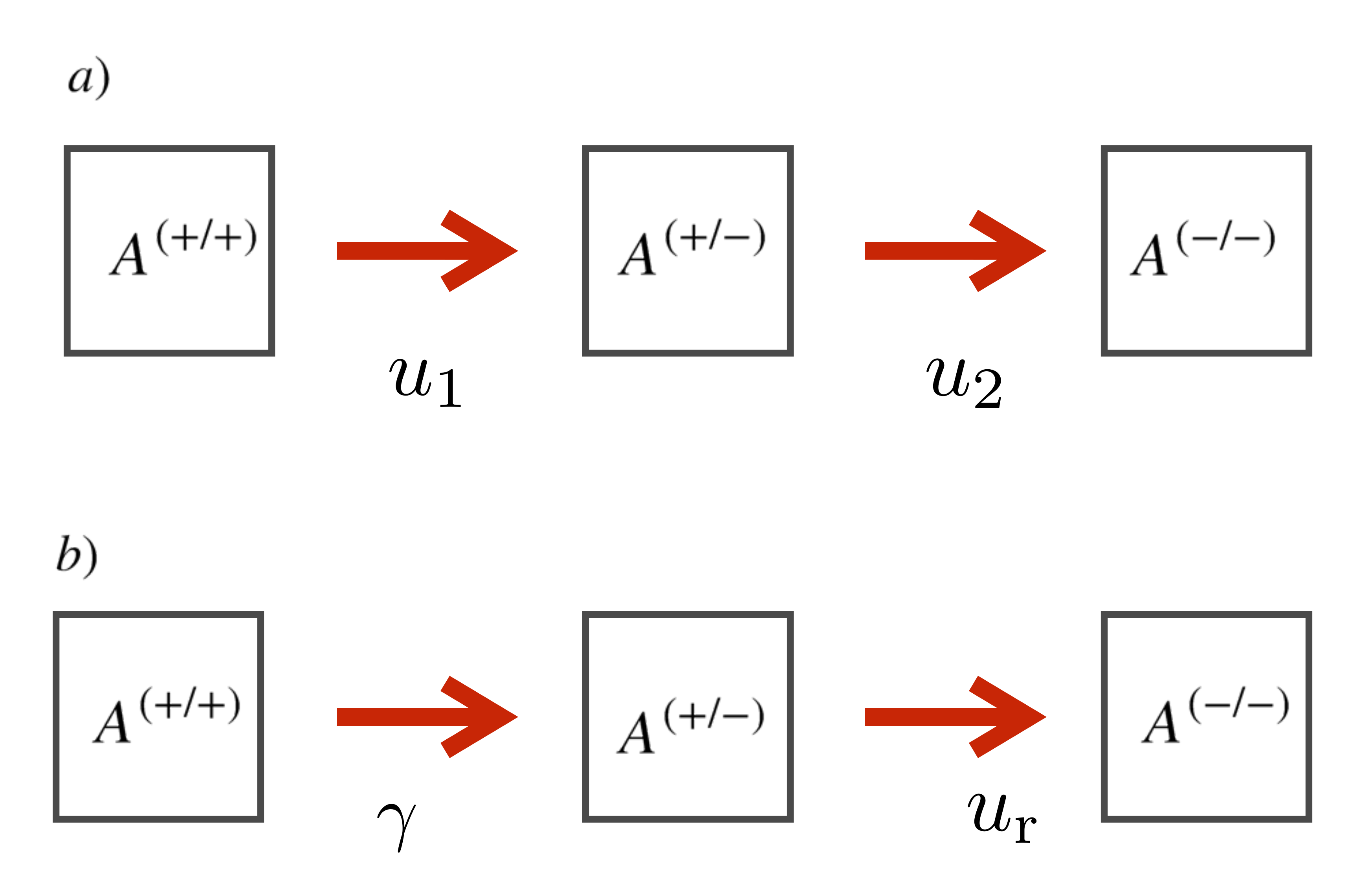, height=170pt,width=255pt,angle=0}
\end{center}
\caption{{\it Two-hit} model of background and radiation-induced second cancer a) In the background two-hits on the wild type allele A, 
occur randomly with rates $u_{1}$ and $u_{2}$. b) For the case of second cancer the first mutation happens during the radiation treatment for a 
small population of normal stem cells (with rate $\gamma$), because of the possible after effects such as genetic instability induced by radiation 
the second cancer mutation rate, $u_{r}$, is assumed to be different from the background rate, $u_{2}$.}
\label{fig1}
\end{figure}

\begin{figure}[!h]
\begin{center}
\epsfig{figure=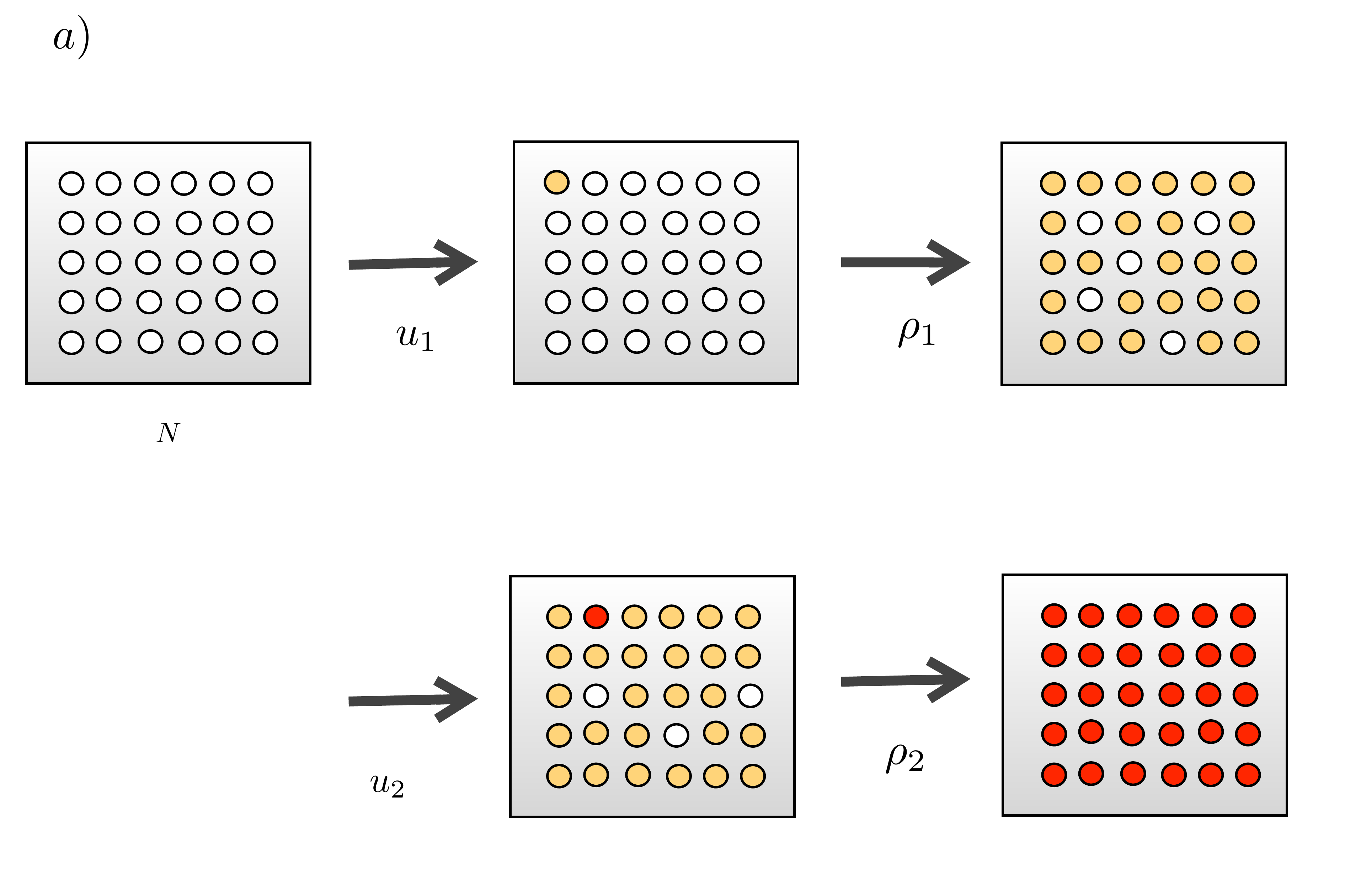, height=200pt,width=300pt,angle=0}
\epsfig{figure=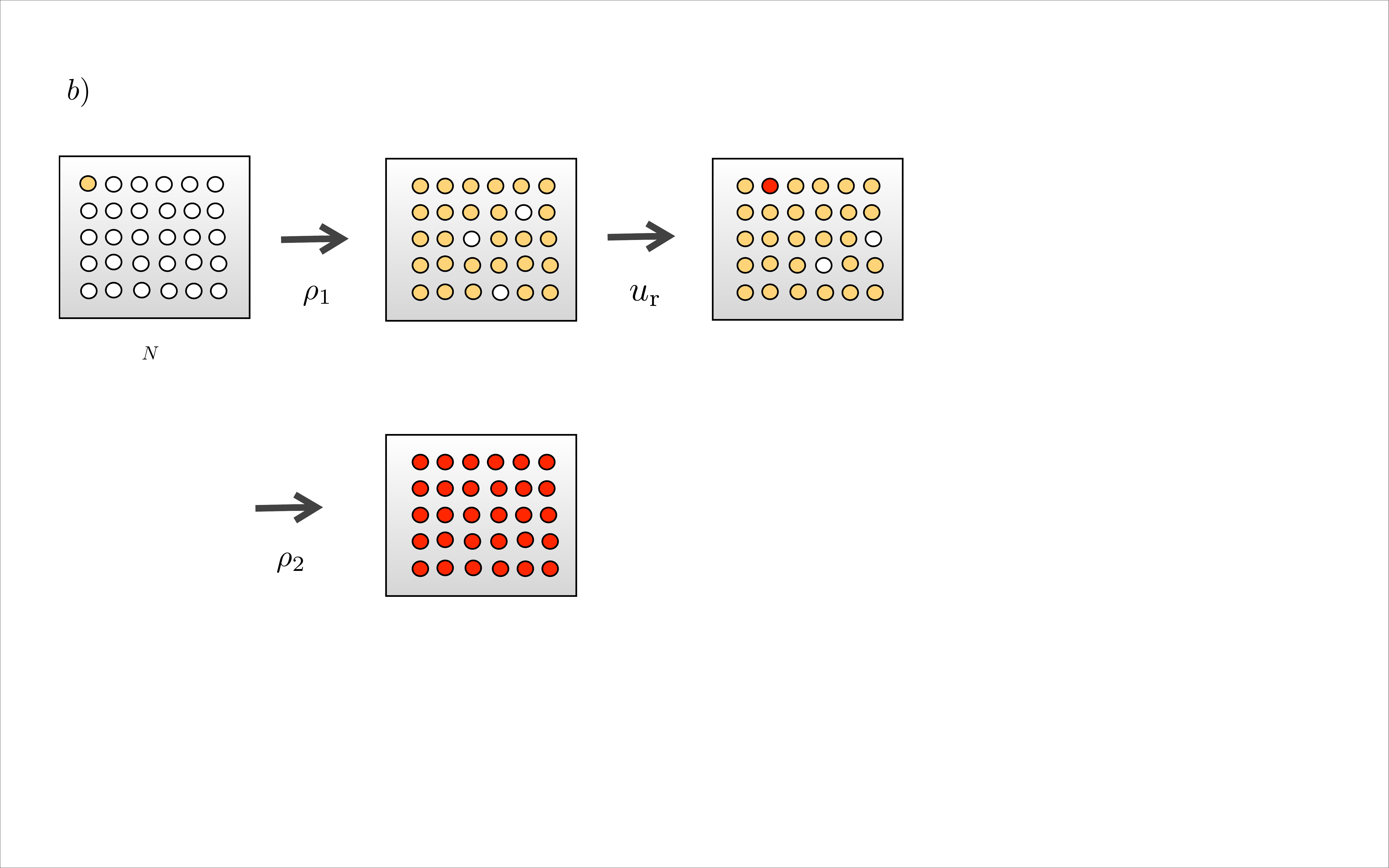, height=200pt,width=300pt,angle=0}
\end{center}
\caption{Normal and pre-malignant cell kinetics until transformation to a malignant niche. a) Background scenario, successive events are a 
first hit mutation (rate $u_{1}$) which creates a pre-malignant cell which outgrows normal cells with probability $\rho_{1}$. Consequently, 
a second hit (rate $u_{2}$) creates a fully malignant phenotype which can outgrow premalignant stem cells with fixation probability $\rho_{2}$. 
b) Post-radiation scenario: A small
number of pre-malignant cells are initiated due to radiation and are distributed among normal stem cell niches, they outgrow in time and can 
transform into malignant cells with a different mutation rate $u_{\rm r}$, (yellow: pre-malignant 
red: malignant).} 
\label{fig23}
\end{figure}

The effects of ionizing radiation in the proposed framework can be summarized as follows:
\begin{itemize}
\item Radiation transforms normal cells into pre-malignant cells. These additional pre-malignant cells are distributed among the niches. 
As stated previously, the contribution of direct two-hit transformations due to radiation is ignored as the event is assumed to be a rare process.
\item There is an effective second hit mutation rate, $u_{r}$ for all the pre-malignant cells that is different from $u_{2}$. The approximate 
value of $u_{\rm r}$ can be estimated from the clinical data for cancer risk estimates.
\item Due to the small numbers of pre-malignant cells relative to the total number of niche areas, only a small subgroup of niches will contain 
pre-malignant cells after treatment.
\end{itemize}
Figure \ref{fig1} depicts a caricature of a two-hit process for background cancer and radiation-induced cancer. The dynamics of transformation 
from normal to pre-malignant tissue is displayed in figure \ref{fig23}.

In the absence of radiation, the probability of having at least one malignant cell (resulting through the sequential mutations $u_{1}, u_{2}$) 
is calculated in a niche before time $t$, i.e. attained age, denoted by $f_{0}(t, u_{1}, u_{2}, r, N)$. For large niche sizes 
(typically $N \geq 10$), a branching process approach is deemed appropriate for the two-hit model.  An explicit form for 
$f_{0}(t, u_{1}, u_{2}, r, N)$ can be obtained (see Appendix A and \citep{key:nowak-popgen}),

\be
f_{0}(t,u_{1},u_{2},r, N) = 1 - \exp\displaystyle\left[-Nu_{1}\cdot \displaystyle\left( \frac{a+b}{c}\cdot \ln\left\{ \frac{e^{ct} + a/b}{1 + a/b}\right\}  - bt \right)\right],
\ee
\noindent
Where coefficients $a, b$ and $c$ are functions of the second mutation rate ($u_{2}$) and proliferation/repopulation rate ($r$) given in 
Appendix A. Upon the appearance of a malignant stem cell in a niche, due to its high fitness advantage, there is a high probability for it to 
take over the whole niche and form a malignant niche. The overall probability of a malignant clone
arising at time $t$ is given by: 
$\rho(\tilde{r},N)f_{0}(t, u_{1}, u_{2}, r, N)$, where $\rho(\tilde{r},N)$ is the fixation probability of malignant cells with a 
proliferation strength $\tilde r$ (the fixation probability is the probability that one malignant cell takes over the whole niche). 
We assume a much higher proliferation rate for malignant cells than for pre-malignant or normal cells and thus the average time to fixation is 
much shorter relative to the time scales of the latency time of secondary cancer, and given by

\be
t_{\rm lag} = \Delta t_{\rm fixation} \sim \frac{1}{r-1}\ln N \ll \frac{1}{u_{2}}.
\label{tlag}
\ee

The left-hand-side of equation  (\ref{tlag}) is the approximate value of the fixation time \citep{key:durrett}. The the fixation time 
for a malignant phenotype is treated as a free parameter that indicates the lag time from the appearance of the first 
malignant stem cell in a niche until it fixates the system.

As a result of radiation exposure, extra numbers of pre-malignant cells are initiated during radiation treatment. This number is on the order of 
one pre-malignant cell per million normal cells (if distributed uniformly in all the niches) \citep{key:sachs}. Hence, it is assumed 
that the radiation-induced as well as the spontaneously created pre-malignant cells are evenly distributed among a subpopulation of niches, 
i.e. there is a small number of stem cells niches which each have one type-1 cell and $N-1$ type-0 cells, soon after radiotherapy, 
while the majority of niches are filled with type-0 cells. The dynamics of these two initial conditions are inherently different. 
The probability of having at least one malignant, type-2, cell arise at time $t$ after radiation exposure (in a niche or compartment), 
is denoted by $f_{1}(t, u_{1}, u_{r}, r, N)$. Notice that the value of the second mutation rate is now changed to $u_{r}$. 
Under a similar branching process approximation, the functional form of $f_{1}(t, u_{1},u_{r},r, N)$ is taken to be:

\be
f_{1}(t, u_{1}, u_{2}. r, N) = \displaystyle \frac{e^{\left\{r(1-u_{2})(a + b)t\right\}}-1}{\displaystyle (1/a)e^{\left\{r(1-u_{2})(a + b)t\right\}} + (1/b)},
\ee

Similarly, the probability of having a malignant niche after time $t$ is \\$\rho(\tilde{r},N)f_{1}(t, u_{1}, u_{2}, r, N)$. 
To estimate the risk of cancer initiation, it is assumed that the probability of cancer initiation is proportional to the number of 
malignant niches. It is also assumed that in both background and radiation-exposed cases the number of malignant cells created before 
the exposure-age is negligible, but there can be a number of pre-malignant cells that underwent spontaneous initiation, and these are denoted 
by $M_{0}$, while the number of radiation induced pre-malignant cells after radiation is denoted by $M_{1}$. Thus, after radiation 
exposure, there are $M_{0} + M_{1}$ niches with one pre-malignant cell distributed per niche; and $\tilde{N} - M_{0}-M_{1}$ niches with zero 
pre-malignant cells. Similarly in the background case the same scenario is assumed, but with only $M_{0}$ niches containing a pre-malignant 
cell. This can be summarized as follows:

\bea
 \begin {minipage}[c]{0,25\textwidth}{\footnotesize number of malignant cells (background)}\end{minipage} &=& N \, \tilde{\rho}\, \{M_{0}\, f_{1}(t, u_{1}, u_{2}, r, N)\nonumber\\ &+&(\tilde{N} - M_{0}) \, f_{0}(t, u_{1}, u_{2}, r, N)\}.\nonumber\\
 \begin {minipage}[c]{0,25\textwidth}{\footnotesize number of malignant cells (irradiated)}\end{minipage}
 &=& N\, \tilde{\rho}\,\{(M_{0} + M_{1})\, f_{1}(t, u_{1}, u_{r}, r, N)\nonumber\\&+&(\tilde{N} - M_{0}-M_{1}) \,
 f_{0}(t, u_{1}, u_{r}, r, N)\},
\label{err}
\eea

\noindent
\nd where $\tilde{\rho} = \rho(\tilde{r},N)$ and $t = t_{a}-t_{e}-t_{\rm lag}$ is the time since exposure, while $t_{a}, t_{e}$ and $t_{\rm lag}$ 
are attained age, age at exposure and lag time as discussed in previous section. The ratio of the above two equations gives a good 
approximation to the ERR for second cancers. Using the definition of Relative Risk (RR) and Excessive Relative Risk (ERR),

\be
RR = 1 + ERR = \displaystyle \left( \begin {minipage}[l]{0,25\textwidth}{\footnotesize number of malignant cells (irradiated)}\end{minipage}\right ) \Bigg / \left(\begin {minipage}[l]{0,25\textwidth}{\footnotesize number of malignant cells (background)}\end{minipage}\right),
\label{err2}
\ee

\nd now, the number of initiated pre-malignant cells during the irradiation time $M_{1}$, can be estimated using the 
initiation-inactivation-proliferation model proposed by Sachs and Brenner \citep{key:sachs}\linebreak (see also \citep{key:shuryak-frac}) 
for a given radiotherapy protocol. $N_{\rm tot} (= N\tilde{N})$ denotes the total number of normal cells, $\gamma$\,(Gy$^{-1}$) denotes the 
initiation rate of normal cells to pre-malignant cells, then a constant dose delivery rate (radiation strength) leads to a simple form for 
$M_{1}$ given by:

\be
M_{1} \approx N_{\rm tot}(e^{\gamma D} - 1),
\ee

\nd
It is again assumed that the number of malignant cells prior to radiation treatment is negligible so that, the number of pre-malignant 
cells $M_0$ that result from spontaneous mutations before the radiation exposure time, $M_{0}$, can be approximated from a simple evolutionary 
Moran model with two populations of normal and pre-malignant cells. This leads to the forms for $M_{1}$ (as in \citep{key:sachs}), and
\be
M_{0} \approx  N ( 1 - e^{-\rho N u_{1} t_{e}}),
\ee

\nd where $t_{e}$ is the age at exposure. Notice that similar forms are used for the incidence rates in the presence and absence of radiation. 
The effects of radiation only appear in the altered value of the second mutation rate with an additional number of pre-malignant cells. The fact 
that the same form equation  (\ref{err}) can be used for the background incidence rate relies on the  assumption that the number of malignant cells 
developed before the radiation exposure time, is negligible. This will lead to a number of new incidences after the radiation exposure age. 
Therefore, these approximations limit the ERR estimates to decades after the radiation exposure. However, this is a well known fact from 
epidemiological studies of second cancers \citep{key:preston}.

\begin{figure}
\begin{center}
\epsfig{figure=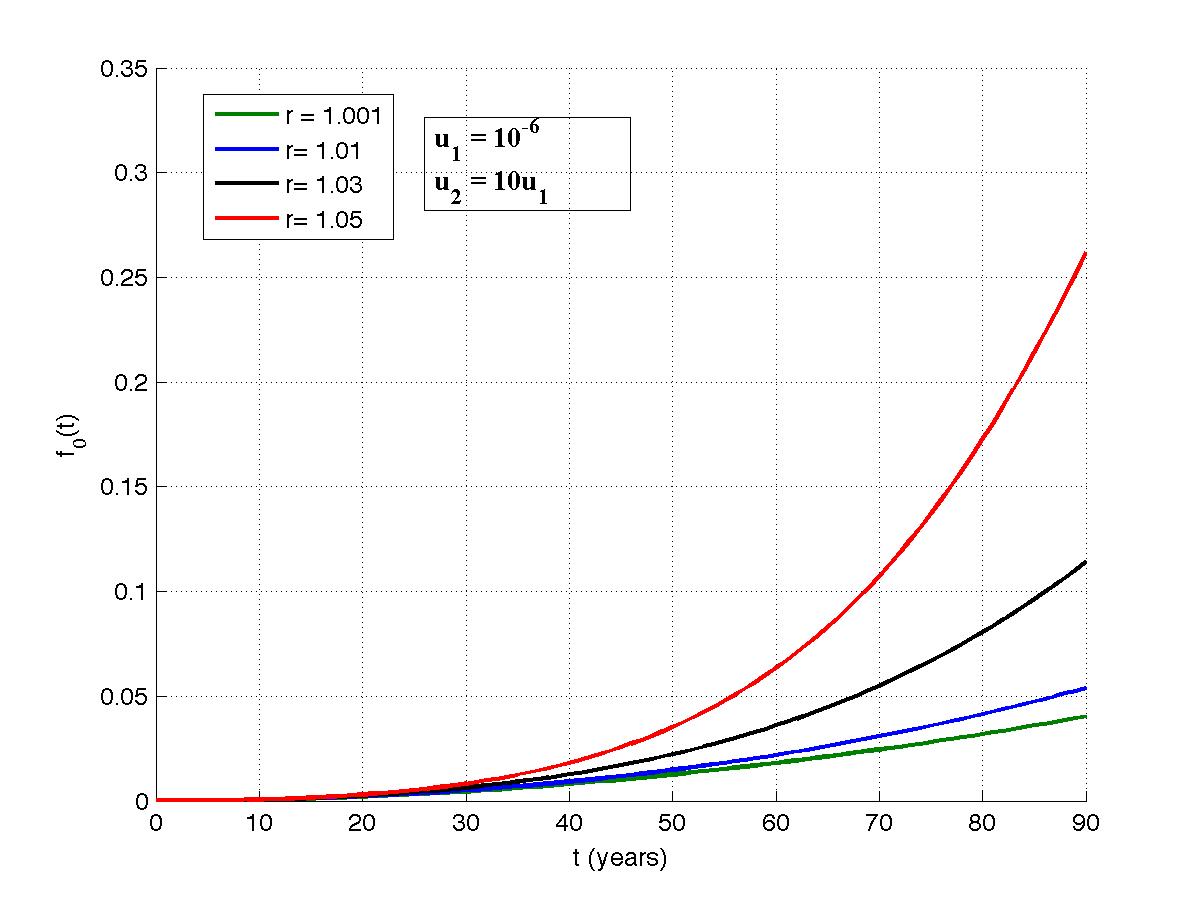, height=220pt,width=240pt,angle=0}
\epsfig{figure=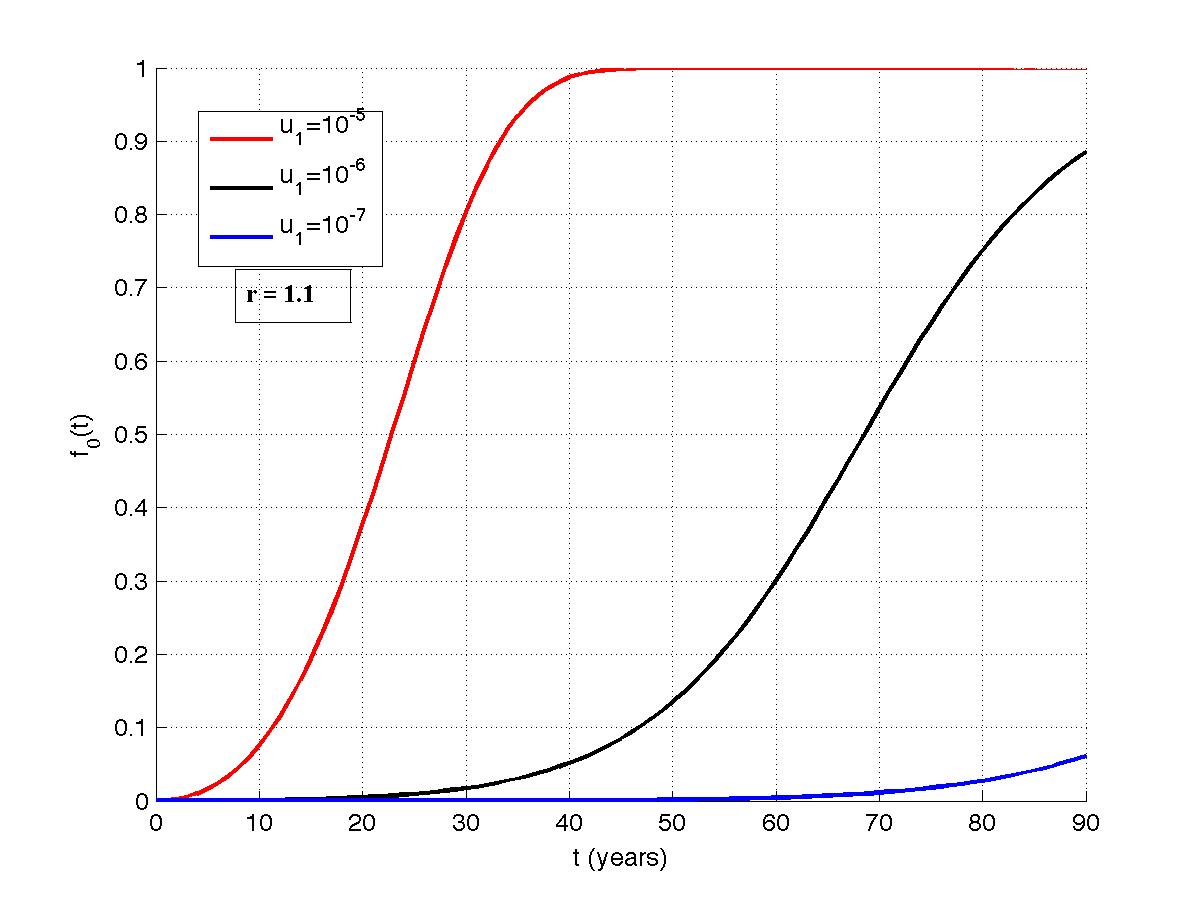, height=220pt,width=240pt,angle=0}
\end{center}
\caption{Probability that a single malignant cell arises (before time $t$) in the system filled with normal cells for various parameter values. 
a) fixed mutation rates and variable proliferation rates $r$. b) Fixed proliferation rate $r$ and variable mutation rates ($u_{2} = 10u_{1}$).}
\label{fig89}
\end{figure}

\begin{figure}
\begin{center}
\epsfig{figure=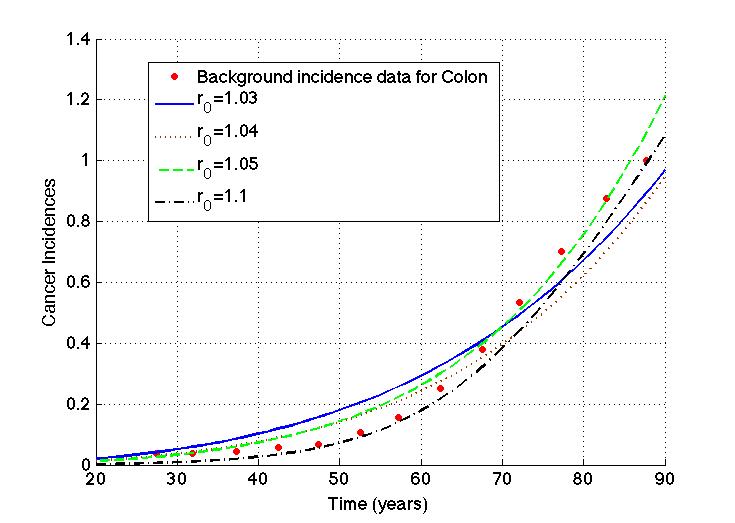, height=170pt,width=190pt,angle=0}
\epsfig{figure=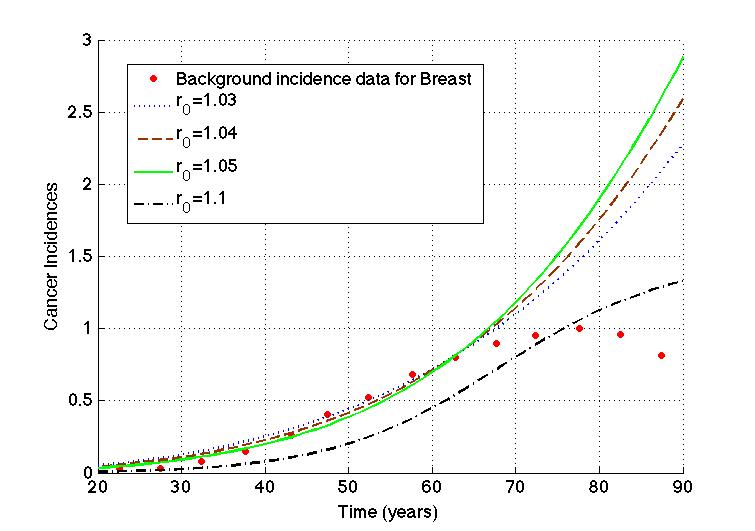, height=170pt,width=190pt,angle=0}
\epsfig{figure=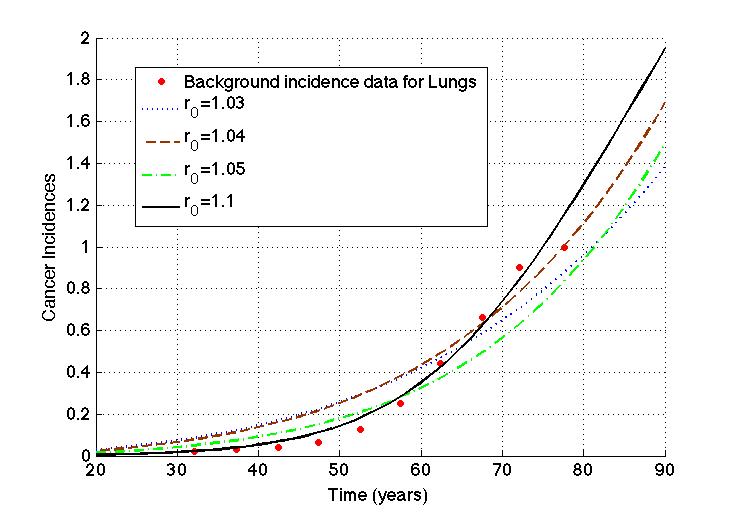, height=170pt,width=190pt,angle=0}
\end{center}
\caption{Model fitting with the background incidence data for colon, lung and breast cancers from SEER database (http://seer.cancer.gov/data/). 
Different graphs correspond to different values of proliferation rate $r_{0}$.}
\label{fig4567}
\end{figure}

Figure \ref{fig89} displays sensitivity of the probability that one malignant cell arises before time $t$ in a system filled with normal 
cells. To show the general behaviour of the function $f_{0}(t, u_{1}, u_{2}, N, r)$ for various parameter values, both changes in mutation rate 
$u_{1}$ and/or proliferation rate $r$ lead to a similar behaviour to that of slowing down the background probability $f_{0}(t, u_{1}, u_{2},N,r)$ 
as a function of time.

\section{Results}
\subsection{Background Age-Incidences}
Using the model described above, (i) the time-dependence of the excess relative risks of radiation for second cancer, and (ii) the 
effective mutation rate of pre-malignant cells resulting from radiation therapy were calculated. The model was fitted with the background 
incidences for colon, breast and lung cancers using data extracted from \citep{key:shuryak1}\citep{key:shuryak2}. The value of the proliferation 
rate, $r$, can be obtained from these fits. In this framework, the niche size has a very weak impact on the value of ERR. It is assumed 
that $N \sim 10$ for the case of colorectal cancer as the stem cell niches are restricted to the bottom of each crypt with small populations. 
For other organs, a much larger value $N \sim 10000$ is used. For the mutation rate values, an estimate from \citep{key:nowak-popgen} 
is employed that approximates the mutation rates for tumor suppressor gene inactivation. A  gene is approximately $10$kb long, and we assume a hit on 5 per cent of bases ($500$ bases) can cause a change-of-function mutation, while each hit might occur with a probability of $10^{-10}$ per cell division \citep{key:nowak-nature}\citep{kunkel}. This gives the 
effective rate of mutation on each gene $u_{1} \sim 10^{-7}$ while each generation is assumed to last about a week. Since the gene can affect either 
of the two alleles, for the second mutation rate, $u_{2}$, the probability of mutation is higher due to mechanisms of loss of heterozygosity 
(mitotic recombination and chromosome non-disjunction). Thus, $u_{2}$ is assumed to be on the order of $10^{-6}$. For the above estimates 
a range of first mutation rates varying between $10^{-8}-10^{-6}$ is considered, while assuming $u_{2}$ to be an order of magnitude larger 
than $u_1$. Using a lag time of $20$ years, results for different cancer incidence models are fitted. This does match the values reported 
in the literature and is also of the the same order of magnitude as that given by equation (\ref{tlag}), which is the theoretical measure of 
the lag time.

The results from Surveillance, Epidemiology and End Results (SEER) database (http://seer.cancer.gov/data/) is plotted in Figure \ref{fig4567} for different values of 
proliferation rates. For the figure, $r = 1.05-1.1$, $N = 10$ for colon cancer; $r=1.03-1.05$, $N=10^{4}$ for breast cancer and 
$r = 1.07-1.1$ with $N=10^{4}$ for lung cancer was chosen.

These values give a crude estimate of the second mutation rate post irradiation. In the following section, an estimate
of $u_{r}$ taken from the atomic bomb survivors data is given. Figure \ref{fig4567} displays the model fit with the background incidence 
data for colon, lung, breast cancers using the above estimated values.

\subsection{Dose-Response in Atomic Bomb Survivors- $u_{r}$ Estimate}

\begin{figure}
\begin{center}
\epsfig{figure=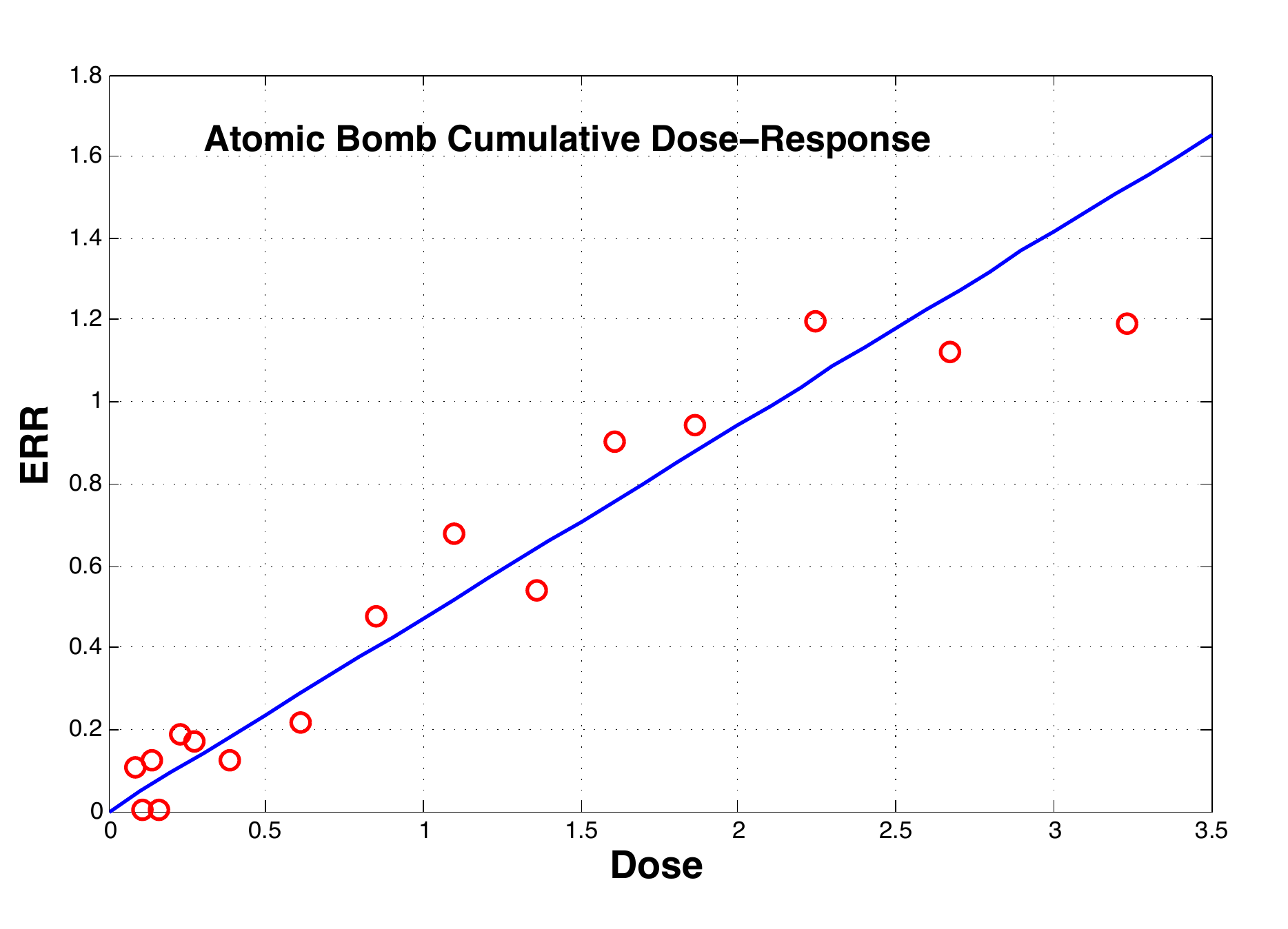, height=200pt,width=250pt,angle=0}
\end{center}
\caption{ERR vs. Dose for total solid cancer incidences, age at exposure of $30$ years and attained age of $70$ years 
(Data points extracted from \citep{key:preston}). The slope of the linear fit is used 
to estimate the mutation rates in a two-stage model of cancer initiation for atomic bomb survivors, (see equation  (\ref{hr-value})).}
\label{fig10}
\end{figure}

The dose response relation for atomic bomb survivors has been reported in the literature (see \citep{key:preston}). The average dose received 
by survivors is about $0$-$3$ Gy \citep{key:preston} (with average estimated to be 0.5Gy). 
The dose-response has a linear functional form for the low dose regions, and gradually bends for high doses (Figure \ref{fig10} displays a plot 
for low dose region - for all solid cancer incidences). The slope of the linear-fit
(which passes through the origin and is indicated by the blue line) is 0.46 per Gy, denotes the weighted absorbed dose, and encompasses the dose
 from neutrons (which is multiplied by an RBE value of 10 to make it equivalent to the gamma dose).

The dose-response relation reported uses the standardized ERR which is for age at exposure of $30$ years and an attained age of $70$ years. 
The dose-response graph can be fitted with a linear function for which the slope indicates the value of ERR per dose. Inserting the value for slope into the analytical form for the ERR, equation  (\ref{err}) and using the rough estimates for 
background mutation rates and proliferation strengths $(u_{1} = 10^{-7}, u_{2}=10^{-6}, r=1.05$ as discussed above), 
we solve for the value of the radiation-induced mutation rate,

\be
u_{r} \simeq 1.0\times10^{-3} ({\rm Gy^{-1}\, generation ^{-1}}).
\label{hr-value}
\ee

Since the dose response plot is a cumulative risk for {\bf all} solid cancers, the above estimate is a good order of magnitude approximation. 
A much higher value of the mutation rate relative to the background value of $u_{2} \sim 10^{-6}$ is noticeable. This is interpreted 
as due to the fact that the secondary effects of radiation probably results in the induction of genomic instability in the model, 
which effectively manifests itself in the value of the second hit probability.

Tables \ref{table1}, \ref{table2} shows the range of values of the second mutation rates as $u_{1}$ is varied between $10^{-8}$ to $10^{-6}$ 
(while keeping $u_{2}\sim 10 u_{1}$ and $r$ varies between 1.001-1.05.

\begin{table}[h]
\begin{tabular}{c}
\includegraphics[width=0.7\textwidth]{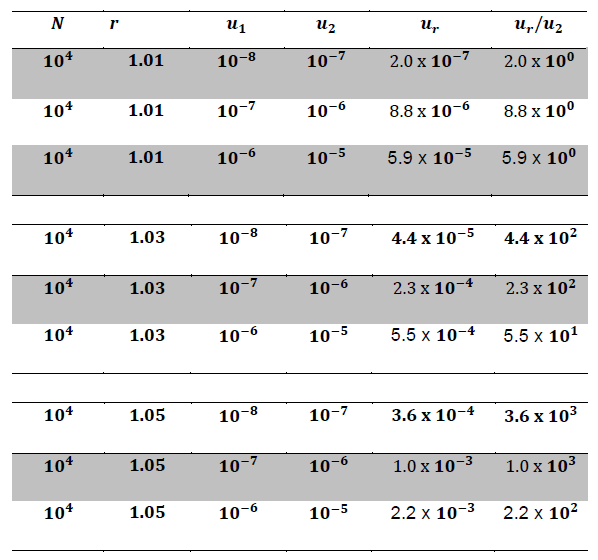}
\end{tabular}
\caption{Best fit mutation rates for background incidence curves, along with the post radiation induced mutation rate.}
\label{table1}
\end{table}

\begin{table}[ht]
\begin{tabular}{c}
\includegraphics[width=0.75\textwidth]{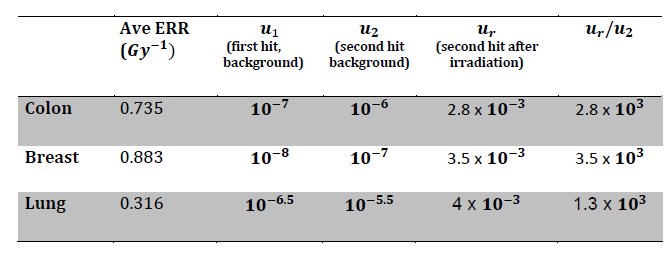}
\end{tabular}
\caption{Second cancer mutation rate estimates for colon, breast and lungs}
\label{table2}
\end{table}

\subsection{ERR vs. Exposure Age, Hodgkin's Lymphoma Survivors}

\begin{figure}
\begin{center}
\epsfig{figure=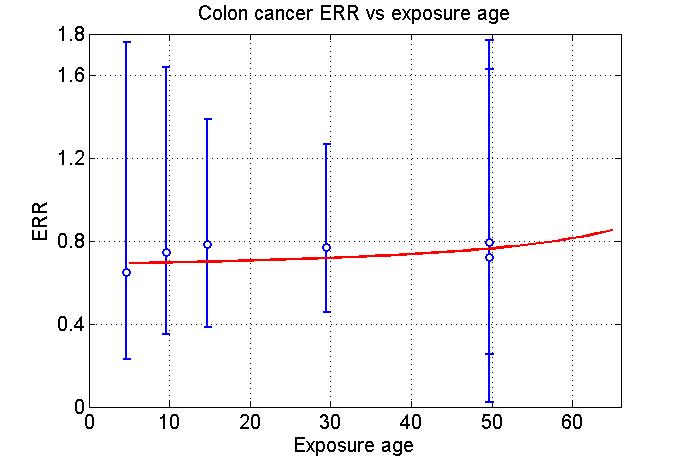, height=180pt,width=250pt,angle=0}
\epsfig{figure=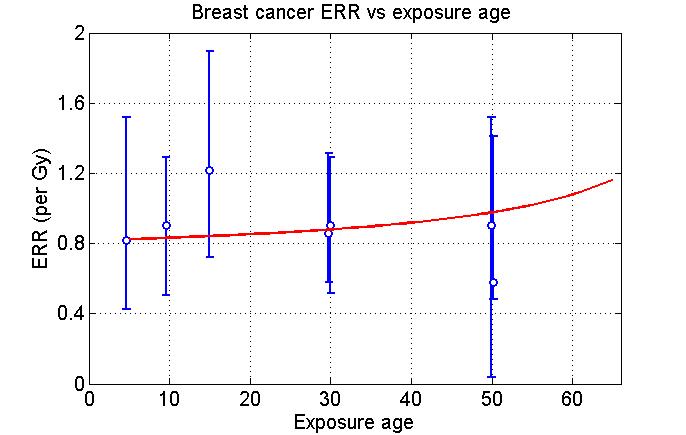, height=180pt,width=250pt,angle=0}
\epsfig{figure=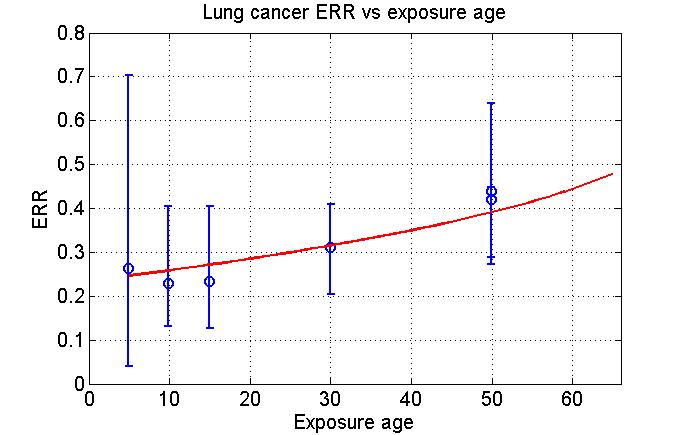, height=180pt,width=250pt,angle=0}
\end{center}
\caption{ERR vs. {\bf exposure} age for colon, breast and lungs among Hodgkins Lymphoma survivors from SEER data (http://seer.cancer.gov/data/ see also \citep{key:shuryak2})
. Solid lines are the best fit from the presented model using background parameter values from 
SEER background incidence rates. }
\label{fig111213}
\end{figure}

In order to obtain similar results for second cancers as a result of radiation therapies, the model is applied to the SEER data on 
second malignancies among Hodgkin's lymphoma survivors. The estimates for the background mutation rates and the proliferation strength 
are further tested by fitting the ERR as a function of exposure age (attained age is fixed at $70$ years). Here, the results
 for three types of cancers, namely, colon, breast and lung, are reported. Using averaged values of ERR over age of exposure for each 
cancer type, we solve for the values of $u_{r}$, and use it in equation  (\ref{err2}) to plot ERR as a function of age at exposure ($t_{e}$). 
Fits were done for values of  $u_{1}$ in the range $\sim 10^{-8}-10^{-6}$, and for small proliferation strengths $r\sim 1.01-1.1$. 
The values of the best fits are reported in Table \ref{table2} and the dependence of exposure age on the ERR is displayed in Figure \ref{fig111213}.

As can be seen, risk is an increasing function of the exposure age as it approaches the attained age of $70$. This is
in fact in agreement with the phenomenological studies of Preston et al. \citep{key:preston} in which a Poisson regression analysis is used 
by assuming a log-quadratic form in their assumption for ERR exposure age dependence.

\subsection{ERR vs. Attained Age- Hodgkin's Lymphoma Survivors}

\begin{figure}
\begin{center}
\epsfig{figure=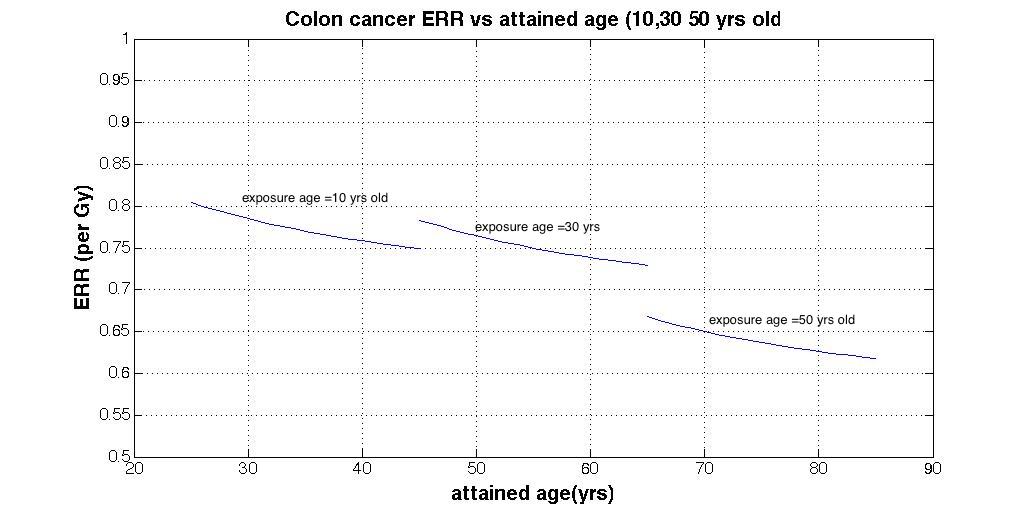, height=190pt,width=300pt,angle=0}
\epsfig{figure=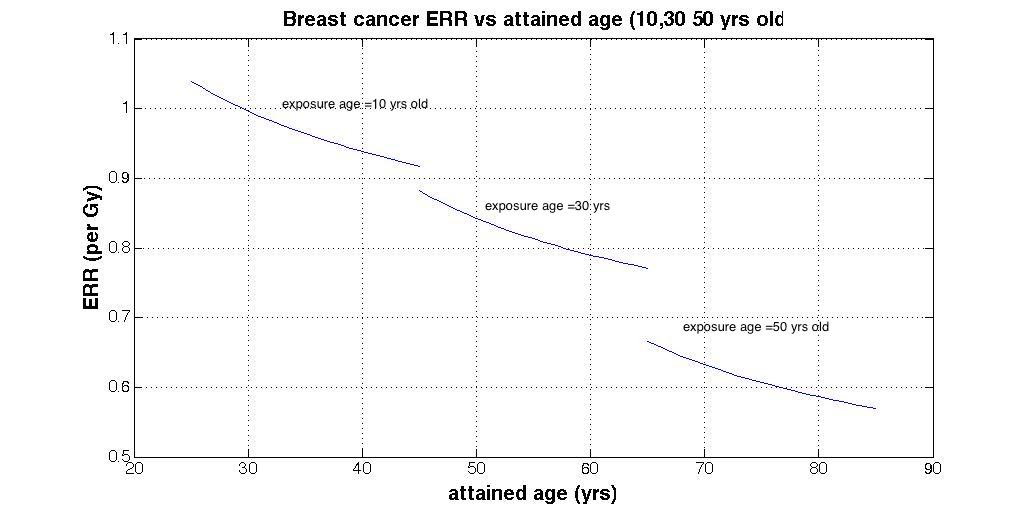, height=190pt,width=300pt,angle=0}
\epsfig{figure=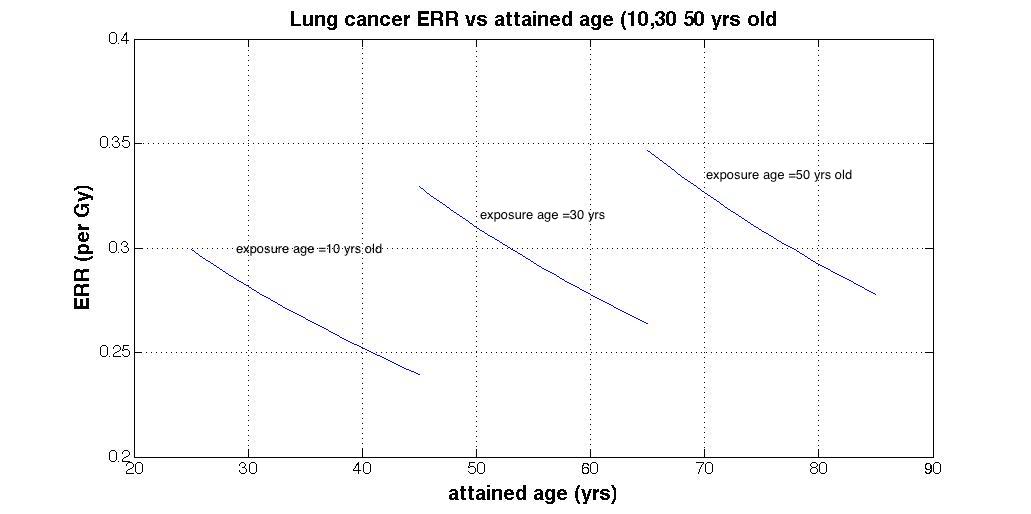, height=190pt,width=300pt,angle=0}
\end{center}
\caption{Predicted values for ERR vs. Attained age for various age groups for colon, breast and lung cancers. The parameters in 
equation  (\ref{err2}) are fixed using SEER background data (http://seer.cancer.gov/data/). Different segments of each graph correspond to 
exposure ages of 10, 30 and 50 years old. A similar distinction is done in \citep{key:preston}(see discussion in the text).}
\label{fig1516}
\end{figure}

The above analysis will allow prediction of the risks of second cancer for different age groups and for different attained ages. The 
predicted ERR's are reported in Figure \ref{fig1516}. Notice that the different age groups of $10$ years, $30$ years and $50$ years display 
different behaviors of the ERR. This is due to the fact that different values of ERR are used as the input to estimate $u_{r}$ for each age group 
in addition to differences in the age at exposure that explicitly appears in equation  (\ref{err2}). In other words, the effect of 
age at exposure on the second cancer risk at a given age (attained age) is two fold: 1) ERR explicitly depends on the age at exposure 
in equation  (\ref{err2}). 2) The value of the second mutation rate, $u_{\rm r}$ depends on the age at exposure and thus should be determined 
separately for each age group. This is done by using the results of ERR vs age at exposure discussed in the previous section. 
While the mutation rate, $u_{\rm r}$, was estimated for standardized attained age of 70 years old in the previous section, exposure ages of 
10, 30 and 50 years old are used from Figure \ref{fig111213} for ERR to estimate a new mutation rate value for each exposure age group. 
Unsurprisingly, this secondary effect is not very significant as the ERR values do not change much for different exposure ages 
(with the exception of lung cancer). The above results can be compared with the phenomenological findings of Preston \citep{key:preston} 
and Shuryak et al  \citep{key:shuryak1}, \citep{key:shuryak2}. The importance of the present result is that one can confirm the 
behavior of the ERR on a more general grounds based on two-stage mutation model, while much of the details of the model such number of 
stem cell niches, $\tilde{N}$, and niche size, $N$, proliferation rate, $r$ and mutation rates $u_{1,2}$ and $u_{\rm r}$ denote the mutation rates.

It is worth mentioning that for organs with strong hormonal production such as thyroid and breast, one does expect a deviation of the 
predicted ERR around certain ages where hormone production can aggravate the risk. This can be seen as a slight deviation in Figure 
\ref{fig111213} (for breast) around an exposure age of 15-20. The reader is referred to \citep{key:shuryak2} for a similar deviation in thyroid 
secondary primary cancers from Hodgkin's Lymphoma survivors. Similar deviations can be observed in the fits for SEER breast cancer incidences 
in Figure \ref{fig4567} after age 60, which 
we propose might be influenced by a similar hormonal mechanism. However, it is suggested that the over all framework especially the estimate 
of second cancer mutation rate and possibility of genomic instability due to radiation is independent of
such details. Further investigation of hormonal effects and its interplay with other mechanisms of radiation induced cancers remains a subject 
for further investigation. 

\section{Discussion}

In the present paper a biologically motivated evolutionary model is described to forecast radiation-induced second cancer risks 
over time, post irradiation. The mathematical framework adopts an evolutionary dynamics approach, particularly with respect to the two-hit 
mechanism used to model cancer initiation. This two-hit process accounts for two types of mutations that occur in normal cells, similar to 
that of tumor suppressor gene inactivation. The assumption is that normal cells can mutate to pre-malignant cells during radiation 
treatment, which in turn transform to malignant cells over a period of time. The second mutation can be correlated to either genomic instability 
(which is considered to be one of the secondary effects of radiation), or, to spontaneous mutation (due to the effects of several carcinogens 
from, for example, food, background radiation, etc).

The current unified framework is an improvement over other models found in the literature. The short-term model allows one to consider 
various  radiation therapy protocols used in current clinical practice. The output of this biological model is then used as an input to 
track the number of malignant niches over time to predict the changes in ERR over time, using a simplified evolutionary model. 
One of the major improvements over existing models is that a stochastic formalism is used to predict ERR over time. This seems a more 
realistic approach for applications to clinical data since mutations appear to occur randomly post irradiation. Another major advantage of 
the presented model is that it has fewer parameters compared to other existing models.

This framework takes into consideration the short-term biological mechanism of pre-malignant cell initiation, along with a long-term 
formalism resulting in the manifestation of a secondary tumor. The short-term model adopts the initiation-inactivation-proliferation model 
of Sachs and Brenner \citep{key:sachs} to estimate the number of radiation-induced pre-malignant cells at the end of treatment. The 
long-term formalism relies on an evolutionary model of the two-hit processes similar to the tumor suppressor gene inactivation model. 
An evolutionary model is developed based on the underlying biological assumptions. The two important parameters in the model are proliferation 
strengths and mutation rates which are estimated from the atomic bomb survivors data. The results from the presented model are consistent 
with those of the work of Preston et al \citep{key:preston}.

In the present paper, the model was applied to patient survivors of Hodgkin's Lymphoma. For this, the proliferation strengths and mutation 
rate parameters were extracted for various organs (in particular: colon, breast and lungs) from the background incidences (taken from the SEER 
database). This was done by fitting the developed analytical functions to the background incidence curves, in order to obtain mutation 
rates and proliferation strengths. Then these values were used to predict ERR's for various attained and exposure ages. The predicted age- 
dependent ERR values are consistent with epidemiological findings and clinical studies. An important point to highlight from the present 
calculations is that the mutation rate post irradiation ($u_r$), is greater than that of the second mutation rate ($u_2$) for background incidence. 
It is also observed that for low proliferation rates and for low risk, the corresponding $u_r$ estimate is similar to $u_2$.

In future work, we intend to extend our framework to include therapeutic doses using dose-volume histograms (DVH) and validate this on existing 
second cancer data sets. It is also intended to extend this work to compare more contemporary protocols for various radiotherapy treatment 
techniques to predict the risk of second cancers for individual patients. There are several ongoing debates on the effects of radiation, 
(e.g concerning genomic instability). It is planned to modify the model to investigate and understand the effects of radiation on 
tissue architecture as well as on genomic instability. \\
\\
{\bf  Acknowledgments}
M Kohandel and S Sivaloganathan are supported by an NSERC/CIHR Collaborative Health Research grant. The authors thank DC Hodgson from 
the Princess Margaret Hospital, Toronto for fruitful discussions.

\begin{appendices}
\numberwithin{equation}{section}
\section{Evolutionary Dynamics of Two-Hit Process}
Consider a population of $N$ cells (inside a niche) governed by the Moran process (type-0,1,2 indicating normal, pre-malignant and malignant 
phenotypes). At each time step a cell is randomly chosen, based on its fitness, to reproduce and another cell is randomly chosen to die. 
In the presence of mutations, such as in the model presented in this paper, at each time step there is an alternative possibility 
(instead of death-birth) that a normal cell may transform into a pre-malignant cell or alternatively a pre-malignant cell may transform into a 
malignant cell. The mutation rate for the first hit is $u_{1}$ and the rate for the second hit (pre-malignant into malignant) 
is $u_{2}$ (or $u_{r}$ after radiation). As discussed in the text, let $f_{0}(t, u_{1},u_{2},r,N)$ be the probability that a first malignant 
cell has emerged {\it before} time $t$ given that the initial population consists of all normal cells in the niche. Similarly, 
let $f_{1}(t, u_{1}, u_{2}, r, N)$ be the probability that a malignant cell emerges in a niche of size $N$ before time $t$, beginning with one 
pre-malignant cell and $N-1$ normal cells. It is straightforward to show that the two probabilities satisfy the continuous time limit of 
the Kolmogorov equations,

\bea
\frac{{\rm d}f_{0}(t)}{{\rm d}t} &=& Nu_{1}f_{1}(t)\left( 1 - f_{0}(t) \right),\nonumber\\
\frac{{\rm d}f_{1}(t)}{{\rm d}t} &=& ru_{2} - (1-r +2ru_{2})f_{1}(t) - r(1-u_{2})f_{1}^{2}(t),
\label{df}
\eea

\nd with the initial condition $f_{0}(0) = 0$. The solutions of the above set of equations can be directly obtained under a branching process 
approximation, i.e. independence of two lineages starting from two different cells, one can then obtain an analytical solution for these
 probabilities,

\begin{eqnarray}
f_{1}(t, u_{1},u_{2},r,N) = \frac{\exp\Bigg\{( r(1-u_{2})(a + b)t\Bigg\}- 1}{(1/a)\exp\Bigg\{ r(1-u_{2})(a + b)t\Bigg\} +(1/b)},\nonumber
\end{eqnarray}

\be
f_{0}(t, u_{1},u_{2},r,N) = 1 - \exp\left( -Nu_{1}\left\{ \frac{a + b}{c}\displaystyle \ln\left[ \frac{e^{ct} + a/b}{1 + a/b}\right] - bt\right\}\right),
\ee

with $a,b,c$ given as,

\bea
a &=& \frac{1}{2(1-u_{2})}\left[ -\left(\frac{1-r}{r} + 2u_{2}\right) + \sqrt{\left(\frac{1-r}{r} + 2u_{2}\right)^{2} + 4u_{2}}~\right],\nonumber\\
b &=& \frac{1}{2(1-u_{2})}\left[ \left(\frac{1-r}{r} + 2u_{2}\right) + \sqrt{\left(\frac{1-r}{r} + 2u_{2}\right)^{2} + 4u_{2}}~\right],\nonumber\\
c &=& r(1-u_{2})(a + b).
\eea

As discussed in the main body of the paper, both the cumulative age-incidence of different cancer types and excess relative risk of cancer 
(second cancer in this case) is expressed in terms of functions $f_{0}(t,u_{1},u_{2},r,N)$ and $f_{1}(t,u_{1},u_{2},r,N)$, \nd which 
lead to a simple form for the ERR. As is obvious there is no dependence on the total number of niches $\tilde{N}$, and the dependence on the 
niche size, $N$, is very weak. However, for more realistic fits the value of the proliferation strength $r$ is not extremely close to unity 
and thus the above approximation might not be appropriate. However, it is easy to see from the above approximate results that any quantity of 
interest (i.e. number of background age-incidences or ERR) is represented by very weak functions of the niche size, $N$. 
Thus the present rough estimates for the value of $N$ do not change any of the important conclusions resulting from this work.

\section{Initiation-Inactivation-Repopulation Model}

In this section the initiation-inactivation-repopulation \linebreak \citep{key:sachs}
formalism is briefly reviewed to estimate the number of pre-malignant cells at the end of radiation treatment. The effect of radiation 
is simplified into two mechanism: 1) cell killing which is modelled by the linear-quadratic approximation and gives the number of either 
normal or pre-malignant cells killed due to radiation dose. 2) The cell initiation which is assumed to be a linear function of the dose with 
a small constant mutation rate per dose rate. Two populations of normal stem cells and pre-malignant stem cells are assumed. Normal stem cells 
grow logistically while they can die during the radiation exposure times, where cell death is given by linear-quadratic formula. They can also 
transform into pre-malignant cells during the exposure time. The population of pre-malignant cells has a similar growth form while its 
proliferation is regulated by the population of normal cells and has the same cell death rate due to radiation and also a positive rate due to 
normal stem cell initiation. The above can be written in the form of the following coupled system of ordinary differential equations.

\bea
\frac{{\rm d}n(t)}{{\rm d}t} = r_{0}n(t)\left(1 - \frac{n(t)}{K}\right) - (\alpha + \beta D)\cdot \frac{{\rm d}D}{{\rm d}t}n(t) -  \gamma\frac{{\rm d}D}{{\rm d}t}n(t),\nonumber\\
\frac{{\rm d}m(t)}{{\rm d}t} = r_{0}\lambda m(t)\left(1 - \frac{n(t)}{K}\right) - (\alpha + \beta D) \cdot \frac{{\rm d}D}{{\rm d}t}m(t) + \gamma\frac{{\rm d}D}{{\rm d}t}n(t),
\eea

\nd where $r_{0}$ is the normal cell repopulation rate while $r_{0}\lambda$ is the pre-malignant repopulation rate. $dD/dt$ is the dose delivery 
rate, $\lambda$ represents the relative proliferation rate of pre-malignant cells to the normal cells and $K$ is the carrying capacity of 
normal cells. $\alpha$ and $\beta$ are the cell killing rates from linear-quadratic dose dependence (LQ). The number of 
pre-malignant cells after the total radiation dose has been applied, is the quantity of importance here. This can be analytically calculated 
for the simplified case of an acute dose, $dD/dt = {\rm const.}$ applied in a finite interval of time

\bea
M &=& m(\infty) = N(\exp(\gamma D) - 1),\nonumber\\
&\sim& N\gamma D,
\label{M-estimate}
\eea

\nd where $D$ is the total dose. The total number of the cells $N_{\rm tot}$, is given by the number of niche times the niches size, 
$N\tilde{N}$. For the values of the radiation-induced initiation rate $\gamma$, this will result in $M/N_{tot} \sim 10^{-6}$ (per Gy). 
Since we are interested on an estimate for number of radiation induced pre-malignant cells, the details of cell killing mechanism in LQ formula 
does not appear in equation  (\ref{M-estimate}).

The above formalism can be expanded to incorporate fractionation therapy, i.e., non-constant dose-delivery rate which has been discussed in the 
literature. In the present work there is no need to deal with these details. However, as a future direction of research all the details of 
radiotherapy (fractionation protocol and the dose-volume histogram for each patient) can be used as input to estimate the ERR as a function of 
exposure age and attained age.

\end{appendices}










\bibliographystyle{spbasic}

\bibliography{mybib-ERR.bib} 

\end{document}